\documentclass[preprint,twocolumn,3p,nopreprintline]{elsarticle}
\usepackage{epsfig}
\usepackage{amsbsy,amssymb} 
\usepackage{graphicx}
\graphicspath{{plots/}} 

\newcommand\new{\newcommand}         

\newenvironment{kinematics}{%
\fontsize{8pt}{10pt}\selectfont%
\begin{tabular}{l|rrrr}
   & \multicolumn{1}{c}{$E$}
   & \multicolumn{1}{c}{$p_x$} 
   & \multicolumn{1}{c}{$p_y$}
   & \multicolumn{1}{c}{$p_z$} \\
\hline}{%
\end{tabular}}

\def\beq{\begin{equation}}   
\def\eeq{\end{equation}}
\def\bea{\begin{eqnarray}}  
\def\eea{\end{eqnarray}} 
\newcommand{\bite}{\begin{itemize}}
\newcommand{\eite}{\end{itemize}}

\def\eps{\epsilon}

\def\eps{\epsilon}

\def\gev{\; \mathrm{GeV}}

\new{\eV}         {{\ifmmode {\mathrm{ eV}}\else ${\mathrm{ eV}}$\fi}}
\new{\MeV}        {{\ifmmode {\mathrm{ MeV}}\else ${\mathrm{ MeV}}$\fi}}
\new{\MeVc}       {{\ifmmode {\mathrm{ MeV}}/c\else ${\mathrm{ MeV}}/c$\fi}}
\new{\MeVcc}      {{\ifmmode {\mathrm{ MeV}}/c^2\else ${\mathrm{ MeV}}/c^2$\fi}}
\new{\GeV}        {{\ifmmode {\mathrm{ GeV}}\else ${\mathrm{ GeV}}$\fi}}
\new{\GeVc}       {{\ifmmode {\mathrm{ GeV}}/c\else ${\mathrm{GeV}}/c$\fi}}
\new{\GeVcc}      {{\ifmmode {\mathrm{ GeV}}/c^2\else ${\mathrm{GeV}}/c^2$\fi}}
\new{\TeV}        {{\ifmmode {\mathrm{ TeV}}\else ${\mathrm{ TeV}}$\fi}}

\new{\Mh}         {{\ifmmode M_{\mathrm{ H}}
                    \else $M_{\mathrm{H}}$\fi}}

\new{\Mz}         {{\ifmmode M_{\mathrm{Z}}
                    \else $M_{\mathrm{Z}}$\fi}}
\new{\Mzsq}       {{\ifmmode M^2_{\mathrm{ Z}}
                    \else $M^2_{\mathrm{Z}}$\fi}}

\new{\as}[1]      {{\ifmmode\alpha^{#1}_s
                    \else$\alpha^{#1}_s$\fi}}
\new{\asx}[1]      {{\ifmmode a^{#1}_s
                    \else $a^{#1}_s$\fi}}
\new{\asb}[1]     {{\ifmmode\overline{\alpha}^{#1}_s
                    \else $\overline{\alpha}^{#1}_s$\fi}}
\new{\asmz}       {{\ifmmode\alpha_s(\Mzsq)
                    \else $\alpha_s(\Mzsq)$\fi}}
\new{\lqcd}       {{\ifmmode\Lambda_{\mathrm{ QCD}}
                    \else $\Lambda_{\mathrm{ QCD}}$\fi}}

\newcommand{\mpi}{Max-Planck-Institut f\"ur Physik, F\"ohringer Ring 6, 80805 M\"unchen, Germany}
\newcommand{\padova}{Dipartimento di Fisica e Astronomia, Universit\`a di Padova, and INFN                                    
Sezione di Padova, via Marzolo 8, 35131 Padova, Italy}
\newcommand{\uiuc}{University of Illinois at Urbana-Champaign, 1110 W Green St, Urbana IL 61801, USA}
\newcommand{\cuny}{New York City College of Technology, City University of New York, 300 Jay Street, Brooklyn NY 11201, USA}
\newcommand{\cunygc}{The Graduate School and University Center, City University of New York, 365 Fifth Avenue, New York, NY 10016, USA}
\newcommand{\cern}{Theory Group, Physics Department, CERN, CH-1211 Geneva 23, Switzerland}

\begin{document}

\begin{frontmatter}


\title{NLO QCD corrections to the production of \boldmath $W^+W^-$ \unboldmath plus two jets\\ at the LHC\tnoteref{t1}}

\tnotetext[t1]{CERN-PH-TH/2012-065, LPN12-040, MPP-2012-48}

\author{N. Greiner}
\ead{greiner@mpp.mpg.de}
\address{\mpi}
\address{\uiuc}

\author{G. Heinrich} 
\ead{gudrun@mpp.mpg.de}
\address{\mpi}

\author{P. Mastrolia}
\ead{pierpaolo.mastrolia@cern.ch}  
\address{\mpi}
\address{\padova}

\author{G. Ossola} 
\ead{GOssola@citytech.cuny.edu}
\address{\cuny}
\address{\cunygc}

\author{T. Reiter} 
\ead{reiterth@mpp.mpg.de}
\address{\mpi}      

\author{F. Tramontano}
\ead{francesco.tramontano@cern.ch}
\address{\cern}

\begin{abstract}
We present the full NLO QCD corrections to the production of a $W^+W^-$ pair 
in association with two jets in hadronic collisions, 
which is an important background to New Physics and Higgs boson searches.  
We include leptonic decays of the W-bosons with full spin correlations. 
We find NLO corrections of the order of 10\% for standard cuts at the LHC.
The scale dependence is considerably reduced by the NLO corrections.
\end{abstract}


\end{frontmatter}



\section{Introduction}

The pair production of electroweak $W$-bosons in association with jets is
an important source  of  
backgrounds for new physics searches at the LHC. 
Events stemming from new physics interactions, like cascade
decays in supersymmetric extensions of the Standard Model,
will produce energetic leptons and multiple jets in combination with missing energy. 
The same signature is produced by Standard Model (SM) processes 
involving the production of leptonically decaying $W$-bosons and jets, where neutrinos escape undetected.
A similar signature also emerges in top-quark pair production within the 
semileptonic decay mode, and in the vector boson fusion (VBF) Higgs-search channels. In the latter case, 
in addition to $W$-pairs originating from Higgs decay, 
the identification of the process relies on the tagging of two forward jets.
Therefore, the direct production of $W^+\, W^- + 2$\,jet  final states constitutes an 
irreducible background to prominent signals. 
Further, they can serve as a test of the electroweak (EW) interactions, allowing to       
measure Higgs boson couplings as well as to probe the presence of anomalous vector boson couplings \cite{Baur:1995uv,Campanario:2010xn}. 

Leading order estimates of the background leave us with a large uncertainty 
about the overall normalisation of the event rates, hence calculations at next-to-leading order (NLO) 
in the strong coupling constant are necessary to obtain reliable predictions of the background
from the SM production of $W^+\, W^-$ plus jets.

NLO calculations of $W$-boson pairs in association with zero or one jet 
showed significant higher order corrections. 
The QCD corrections to $pp\rightarrow W^+W^-$ have been calculated in
Refs.~\cite{Ohnemus:1991kk,Frixione:1993yp}, while the EW corrections have been computed in~\cite{Accomando:2004de}. 
Phenomenological studies 
can be found e.g. in~\cite{Campbell:1999ah,Dixon:1999di,Grazzini:2005vw,Campbell:2009kg,Campbell:2011bn}. 
NLO QCD corrections to $W^+W^-$ production within the Randall-Sundrum model have been calculated in \cite{Agarwal:2010sn}.
$pp\rightarrow W^+W^-$ plus one jet at NLO, 
including decays to leptons, has been studied in
Refs.~\cite{Campbell:2007ev,Dittmaier:2009un,Bern:2008ef}, where a K-factor of about 1.3 has been found. 
Further, the loop induced process 
$gg \to W^+W^-$ including leptonic decays has been calculated in Refs.~\cite{Binoth:2005ua,Binoth:2006mf,Campbell:2011cu}, 
for phenomenological studies with emphasis on Higgs search see also \cite{Duhrssen:2005bz,Campbell:2011bn}.
It was found that the application of realistic Higgs search selection cuts
increases the correction from this channel up to 30\% of the total 
$pp\rightarrow W^+W^-\to \, leptons$ cross section.

A $W$-boson pair in association with two jets can be produced through both vector boson fusion (VBF) 
or production mechanisms without weak bosons in the $t$-channel.
NLO QCD corrections to the VBF mechanism have been calculated in \cite{Jager:2006zc,Jager:2009xx}.
More recently, NLO QCD calculations have been presented for the non-VBF case:
The production of equal-charge $W$-bosons, $W^+W^+jj$, has been computed at NLO accuracy in \cite{Melia:2010bm}. 
It also has been combined with a parton shower\,\cite{Melia:2011gk} within the {\tt POWHEG} framework 
\cite{Nason:2004rx,Frixione:2007vw}, and later has been complemented by EW corrections \cite{Jager:2011ms}.
First results for NLO QCD corrections to the process $W^+W^- jj$ at hadron colliders have been presented in \cite{Melia:2011dw}, see also \cite{Campanario:2011cs}. 

In this letter we present an independent calculation of 
$p p \to W^+W^- j\,j$, which takes into account 
contributions previously omitted. In particular, we also consider diagrams where the 
$W^+W^-$ pair or the $Z/\gamma$ bosons couple directly to a virtual quark loop and the contributions 
from the third quark generation in closed fermion loops, and quantify their effects on the total cross section.
We also include the decay of the $W$ bosons into leptons retaining
full spin correlation exploiting the narrow width approximation.

\vspace*{3mm}

In the past years, very rapid progress in the calculation of 
higher order corrections to multi-particle processes 
has been achieved, both due to the emergence of unitarity based methods 
and the further development of techniques using a tensor  
reduction of Feynman diagrams.
As a consequence, NLO results for hadron collider processes 
with up to five particles in the final state have been 
produced in the last few years, for processes involving 
one-loop functions with six or more external legs see e.g. 
$pp\rightarrow W/Z  + 4$\,jets \cite{Berger:2010zx,Ita:2011wn}, $pp\to 4$\,jets \cite{Bern:2011ep}, 
$pp\rightarrow W (Z,\gamma) + 3$\,jets \cite{Berger:2009ep,Berger:2009zg,KeithEllis:2009bu,Melnikov:2009wh,Berger:2010vm,Campbell:2010cz},
$pp \rightarrow t\bar{t} b \bar{b}$ \cite{Bredenstein:2009aj,
  Bredenstein:2010rs,Bevilacqua:2009zn}, $pp \rightarrow t \bar{t} +
2$ jets \cite{Bevilacqua:2010ve}, $pp \rightarrow b \bar{b} b \bar{b}$\,\cite{Binoth:2009rv,Greiner:2011mp}, $p p \rightarrow t\bar{t} \rightarrow W^+W^-b
\bar{b}$ \cite{Bevilacqua:2010qb,Denner:2010jp}, $pp \rightarrow
W^+W^+ + 2$ jets \cite{Melia:2010bm}, $pp \rightarrow W^+W^- + 2$ jets\,\cite{Melia:2011dw}, 
$W\gamma\gamma+$\,jet\,\cite{Campanario:2011ud},
$e^+e^- \rightarrow \,\geq 5$\,jets \cite{Frederix:2010ne,Becker:2011vg}.

The evaluation of the NLO QCD corrections for $W$-pair production in association with two jets 
can also be considered an important benchmark in view of the advances in the development of packages 
for fully automated NLO calculations\,\cite{vanHameren:2009dr, Hirschi:2011pa, Bevilacqua:2011xh,
Mastrolia:2010nb, Cullen:2011ac}.
As described in the next section, for our calculation we employ the combined use of  
{\tt MadGraph}~\cite{Stelzer:1994ta}, {\tt MadDipole}~\cite{Frederix:2008hu,Frederix:2010cj},
{\tt MadEvent}~\cite{Maltoni:2002qb}, and {\sc GoSam}~\cite{Cullen:2011ac}.
Our results demonstrate the ability of the {\sc GoSam} package to handle the evaluation of highly non-trivial 
one-loop amplitudes in an efficient and automated way. 
In particular, the possibility to filter certain classes of diagrams and process them separately allows us to
quantify the relative impact of different contributions.
For example, we find that the contribution from quarks of the 3rd generation in the loops amounts to 
about 3\% of the virtual contributions restricted to two generations. 
Overall, the NLO corrections at $\mu=2\,M_W$ increase the cross section by about 10\%, and the scale dependence 
is considerably reduced.

This paper is organized as follows. In Section \ref{sec:calc} we describe technical details of the calculation, while
Section \ref{sec:res} contains some phenomenological studies. 
Finally, in Appendix A we provide numerical results                     
for the virtual amplitudes of the different partonic sub-processes                    
for comparison.


\section{Technical Details}
\label{sec:calc}

\subsection{Description of the Calculation}

The leading order (LO) and the real radiation matrix elements are generated using {\tt MadGraph}~\cite{Stelzer:1994ta,Maltoni:2002qb,Alwall:2007st}.
For the subtraction of the infrared singularities we use Catani-Seymour dipoles~\cite{Catani:1996vz},
supplemented with a slicing parameter $\alpha$ as proposed in~\cite{Nagy:1998bb}
implemented in the package {\tt MadDipole}~\cite{Frederix:2008hu,Frederix:2010cj}.

The code for the evaluation of the virtual corrections is generated by the program 
package {\sc GoSam}~\cite{Cullen:2011ac} which combines the automated approach 
of diagram generation and processing~\cite{Reiter:2009ts,Cullen:2010jv} with 
integrand-reduction 
techniques~\cite{Ossola:2006us,Ossola:2007bb,Ellis:2007br,Ossola:2008xq,Mastrolia:2008jb,Mastrolia:2010nb,Heinrich:2010ax}
supplemented by the automatic algebraic computation of the full rational part.

For the tree-level, the real emission part and the integrated subtraction terms, the integration over the phase space 
is carried out using an in house version of {\tt MadEvent}~\cite{Maltoni:2002qb}. We have used $2\cdot 10^7$ phase
space points for the tree-level integration and the integrated subtraction terms, and $3\cdot 10^7$ for
the real emission part.
For the the virtual contributions
we have produced several samples of unweighted events according to the tree-level processes and 
performed a reweighting of the events to obtain the virtual corrections. This has been done independently
for each subprocess. In total, which means summed over all subprocesses, we have used $~4\cdot10^5$ unweighted events.
In order to parallelize as much as possible the reweighting has been done separately
for each non-vanishing helicity.
The unweighted events also have been produced with {\tt MadEvent}.

\subsection{Partonic Subprocesses}

The description of LO and NLO corrections to $p p \to W^+ W^- j j$ requires two different partonic structures, namely
processes with either four quarks or two quarks and two gluons, in addition to the $W$-pair.


The cross section for the associated production of a $W$-boson pair               
accompanied by $b$-jets is dominated by  $t\bar{t}$
on-shell production and decays. These processes have been studied in~\cite{Bevilacqua:2010qb,Denner:2010jp,Melnikov:2011ai}.    
Here we assume the use of an efficient anti-$b$-tagging to exclude those          
processes and concentrate on the                                                
``genuine", i.e. not top-mediated production of the two $W$-bosons and two light jets.
In this sense we work in the four flavour\,scheme and treat the bottom
quarks as massive particles.
We use four flavour pdfs and no mixing among the generations.
The bottom and top quark contributions to the gluon self-energy are
subtracted at zero momentum
transfer and the running of $\alpha_s$ is given by the four light flavours only.
Furthermore, we only take into account doubly-resonant diagrams, i.e. diagrams
with two $W$-propagators. The effect of singly-resonant diagrams, with only one $W$-boson
radiated off an intermediate lepton stemming from a $\gamma$/$Z$-propagator,
is strongly suppressed by phase space, as has been already noted in \cite{Dittmaier:2009un}.
We do not include any Higgs exchange diagrams nor interferences with electroweak production mechanisms of 
$WW\,jj$ final states.

We used the {\sc GoSam} package to automatically produce the amplitudes, and chose dimensional reduction as 
regularisation scheme. 
$\overline{MS}$ renormalisation supplemented by on-shell subtraction for the massive quarks is also done automatically 
by {\sc GoSam}. 

The eight basic partonic processes listed below are the building blocks  
from which all the other subprocesses can be obtained via crossings and/or change of quark generation.
We assume the $W$-bosons decay as $W^+ \to \nu_{e} \, e^+$ and $W^- \to \bar{\nu}_{\mu} \, \mu^- $, 
but as we use massless leptons, the final results for other lepton flavour/charge assignments 
are identical.
Six of these subprocesses do not involve gluons in initial or final states, namely
\begin{eqnarray*}
d \, \bar{u} & \to & c \, \bar{s} \, W^+\, W^- \\
u \, \bar{u}   &  \to & c \, \bar{c} \, W^+\, W^- \\
d \, \bar{d} & \to & s \, \bar{s} \, W^+\, W^- \\
u \, \bar{u}   &  \to & d \, \bar{d} \, W^+\, W^- \\
u \, \bar{u}   &  \to & u \, \bar{u} \, W^+\, W^- \\
d \, \bar{d} & \to & d \, \bar{d} \, W^+\, W^- 
\end{eqnarray*}
while other two subprocesses also involve two gluons:
\begin{eqnarray*}
u \, \bar{u} & \to & g \, g \, W^+\, W^- \\
d \, \bar{d}   &  \to & g \, g \, W^+\, W^- 
\end{eqnarray*}
In Appendix~A we give numerical results for each subprocess, 
labeling each according to the involved partons, i.e. with
$( d, \bar{d}, g, g)$ we mean the subprocess $d \, \bar{d} \to  g \, g \, W^+(\nu_{e} \, e^+)\, W^-(\bar{\nu}_{\mu} \, \mu^-)$.

In the calculation of the virtual corrections for each of the subprocesses, the Feynman diagrams have been divided into 
three groups (called A, B, and C) which we process separately.
Group A is defined by neglecting all contributions from the quarks of the third generation in the fermion loops
and neglecting contributions in which vector bosons are attached to closed fermion loops,
which is the approximation used in Ref.~\cite{Melia:2011dw}. 
Group B contains only diagrams with vector bosons attached to closed fermion loops, and group C contains all contributions from the 
third generation quarks in the loops. 
The last two sets of diagrams, which are found to be numerically less significant, have not been
considered previously in the  literature.

\subsection{Numerical Checks}

The leading order and the real emission matrix elements, which are generated by 
{\tt Madgraph}, have been compared with the corresponding matrix elements generated by {\sc GoSam}.
The infrared singularities computed with {\tt MadDipole} have been checked with dedicated routines implemented in {\sc GoSam}. 

Concerning the virtual amplitudes, we have reproduced the results of known processes 
with a similar structure as the one presented here.
We re-computed $u \, \bar{u} \to b \, \bar{b}\, W^+\, W^-$ and  $g \, g \to  b \, \bar{b}\, W^+\, W^-$  
with on-shell $W$-bosons and massless $b$-quarks~\cite{vanHameren:2009dr} 
starting with off-shell $W$'s to recover the on-shell results from 
(multiple-evaluation of) the leptonic $W$-decay modes. The process $u \, \bar{u} \to b \, \bar{b}\, W^+\, W^-$ was also checked 
considering the $W$'s directly as on-shell particles.

In the case of decaying $W$-bosons, we also checked all results presented in \cite{Melia:2010bm} for $pp \rightarrow
W^+W^+ + 2$ jets and in \cite{Melia:2011dw} for $pp \rightarrow W^+W^- + 2$ jets. 
Further details about this last comparison are provided in the Appendix.


In addition, we have  performed several checks on the integrated results. 
For the tree-level and the real emission matrix element, we have compared two different ways of obtaining 
the same cross section. 
We have treated the $W$-bosons as stable particles, which leads
to a reduction of the phase space and hence to an easier integration. We have then multiplied this result by the branching
fractions to leptons and compared it with the full result including the $W$ decays where no cuts on the leptons have been
applied. This check has been performed for the most complicated class of subprocesses involving two quarks and two gluons
and we have found agreement of the two results within the statistical uncertainty.

As a test on the subtraction terms, we have checked the independence of the ``slicing'' parameter $\alpha$ 
for several classes of subprocesses.

The procedure of reweighting for the virtual contributions has been checked by a comparison of the
total cross section with a result obtained using a full phase space integration. 
As these checks are very CPU-intensive,  we restricted ourselves to two classes of subprocesses, 
namely the case with four different quark flavours 
and the subprocess with two $u$-type quarks and two gluons, 
the latter providing the largest contribution to the total cross section.
Both results agree within the statistical error, which indicates that the phase space is fully resolved
by the integration routine. 
For our final results, about one permil of the phase space points have been classified as numerically unstable and 
discarded.


\section{Phenomenological Results}
\label{sec:res}

In this
section we present a selection of phenomenological results 
for proton proton collisions at the LHC at $7$\,TeV.
A complete phenomeno\-logi\-cal analysis and a more detailed study of all the sources of 
uncertainties is beyond the scope of the present paper and therefore left  
for a future publication.

\subsection{Setup}
For all the results and distributions shown in this section we have used the
parameters listed below.
\begin{center} 
{\small
\begin{tabular}{|l|l|}
\hline
\multicolumn{2}{|c|}{Parameters}\\
\hline
$M_W = 80.399 \gev$ & $\Gamma_W = 2.085 \gev$ \\
$M_Z = 91.188 \gev$ & $\Gamma_Z = 2.4952 \gev$ \\
$M_t = 171.2 \gev$ & $\Gamma_t = 0. \gev$ \\
$M_b = 4.7 \gev$ & $\Gamma_b = 0. \gev$ \\ 
$\alpha(M_Z) = 1/128.802$  & $c_W^2 = M_W^2/M_Z^2 $ \\ 
\hline
\end{tabular} 
 }
 \end{center}

The weak mixing angle is calculated from the $W$ and $Z$ mass.
The strong coupling constant and its running are determined by the set of parton distribution functions. 
We used the MSTW2008 pdf set \cite{Martin:2009iq}, where the values for $\alpha_s$ at leading order 
and next-to-leading order are given by
$$
\alpha_{s,LO}(M_Z) = 0.13355\;, \; \alpha_{s,NLO}(M_Z) = 0.1149\;, 
$$
and the running is calculated at one loop for the tree-level result and at two loops for the next-to-leading order
 parts. As we neglect initial state $b$-quarks, we use the $N_f=4$ version of the pdf set.\\
For the  jet clustering we used an anti-$k_T$ algorithm with a cone size
of $R=0.4$ provided by 
the FastJet package \cite{Cacciari:2005hq,Cacciari:2008gp}.

\subsection{Cuts}
We show results for the LHC at $\sqrt{s}=7$\,TeV, using a set of standard LHC cuts given by
\beq
p_{T,j} \ge 20  \gev, \quad |\eta_j| \le 3.2, \quad \Delta R_{jj} \ge 0.4
\eeq
for the jets and
\beq
p_{T,l} \ge 20  \gev, \qquad |\eta_j| \le 2.4
\eeq
for the charged leptons. In addition we impose a cut on the missing transverse energy of
\beq
E_{T,miss} \ge 30 \gev.
\eeq
Neglecting final state $b$-quarks is only justified if the $b$-jets can be 
(anti-)tagged. As $b$-tagging only works efficiently if the jets are produced centrally, we  do not allow
for jets with a very large rapidity.

\subsection{Results}
\label{sec:dist}
\begin{figure}[htb]
\begin{center}
\includegraphics[width=7.5cm]{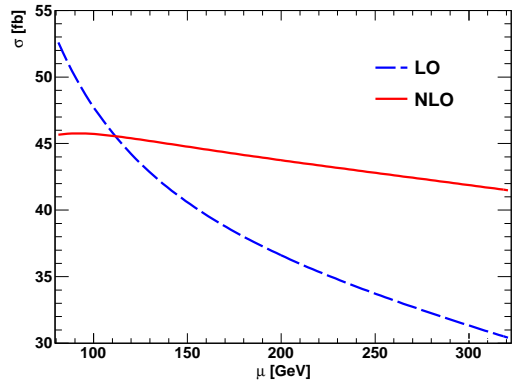}
\caption{Dependence 
of the LO and NLO cross sections 
$pp\rightarrow  
W^+(\rightarrow\nu_e \,e^+) W^-(\rightarrow \mu^- \bar{\nu}_\mu)\, j j$ at $\sqrt{s}=7$\,TeV
on renormalisation and factorisation scales. 
We use $\mu=\mu_R=\mu_F$ and vary between $M_W\leq \mu\leq 4\,M_W$. }
 \label{fig:scalevar}
\end{center}
\end{figure}

In Fig.~\ref{fig:scalevar} we show the dependence of the total cross section on the
variation of the renormalisation and factorisation scales.
The cross section, within the cuts given above, has been evaluated
for the three values of $\mu_R=\mu_F=M_W,2\,M_W,\,4\,M_W$. 
As expected, adding the next-to-leading order
corrections strongly reduces the theoretical uncertainties.
Choosing the scale $\mu=2\,M_W$ as the central scale, we find \\
%
$\sigma_{\rm LO}[fb]=39.57 {\strut+34\%\atop-23\%} (\mathrm{scale}) \pm 0.13\% (\mathrm{stat.}) $,\\
$\sigma_{\rm NLO}[fb]=44.51 {\strut+2.5\%\atop-7.4\%} (\mathrm{scale}) \pm 0.70\% (\mathrm{stat.}) $,\\
%
where we can see a clear reduction of the scale dependence at NLO. 
The statistical error coming from the numerical integration is negligible compared to the residual scale uncertainty.

In the distributions the reduction of the theoretical uncertainties also becomes apparent.
The shaded areas in the distributions denote the change of the distribution when varying the
scales  $M_W \le \mu_R=\mu_F \le 4\cdot M_W$. All the distributions also include estimates
of the pdf uncertainties. To estimate the pdf uncertainty we calculate the variables
$\Delta X_{max}^{+}$ and $\Delta X_{max}^{-}$ which can be derived from the set of eigenvectors
that are included in the pdf set. The exact definition of these variables and a detailed discussion
can be found in \cite{Alekhin:2011sk}. For this calculation we used the eigenvectors of the
$90$ percent confidence level set. Each phase space point enters the plots three times.
Once with the central value of the pdf set and in addition for the two variables
$\Delta X_{max}^{+}$ and $\Delta X_{max}^{-}$. However for the actual phase space integration,
i.e. in the calculation of the total cross section and the adaption of the integration grid
during runtime, only the central value of the pdf set enters. In this way we are able to
include an estimation of the pdf error into the histograms. 

\begin{figure}[htb]
\begin{center}
\includegraphics[width=7.5cm]{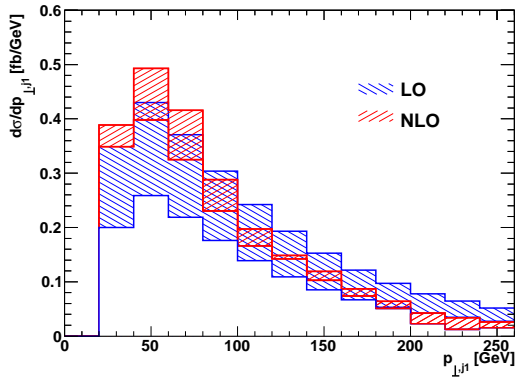} 
\includegraphics[width=7.5cm]{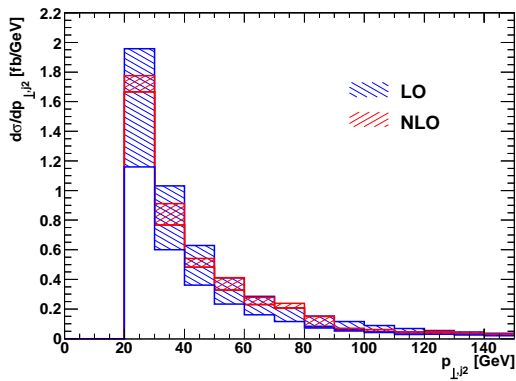}
\caption{Transverse momenta of the two tagging jets
at $\sqrt{s}=7$\,TeV.
The jets are ordered according to their $p_\perp$.
The bands describe the pdf and statistical error as well as the scale variations
between $M_W\leq \mu\leq 4\,M_W$.\label{fig:pt} }
\end{center}
\end{figure}

\begin{figure}[htb]
\begin{center}
\includegraphics[width=7.5cm]{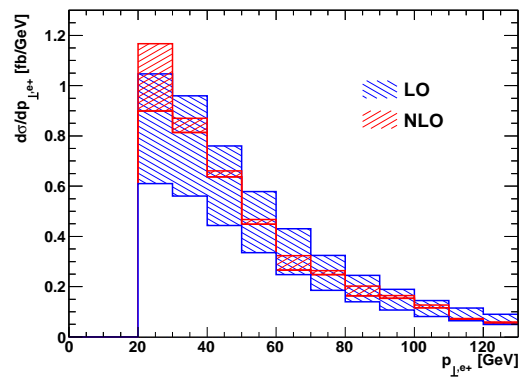} 
\caption{Transverse momentum of the electron, in 
$p p \rightarrow 
e^+\nu_e\mu^-\bar{\nu}_\mu\,j j$
  at $\sqrt{s} = 7$\,TeV. }
\label{fig:pte}
\end{center}
\end{figure}

\begin{figure}[htb]
\begin{center}
\includegraphics[width=7.5cm]{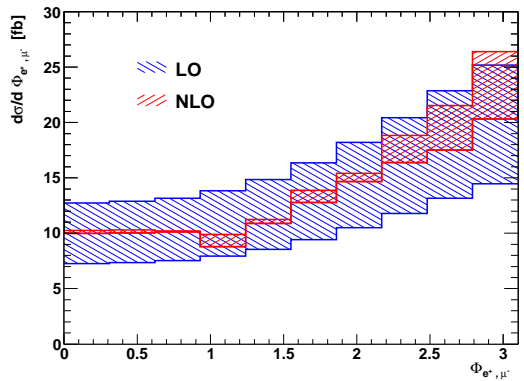} 
\caption{Distribution for the opening angle between the
  leptons, $\phi_{e^+\mu^-}$.}
\label{fig:phi}
\end{center}
\end{figure}

\begin{figure}[htb]
\begin{center}
\includegraphics[width=7.5cm]{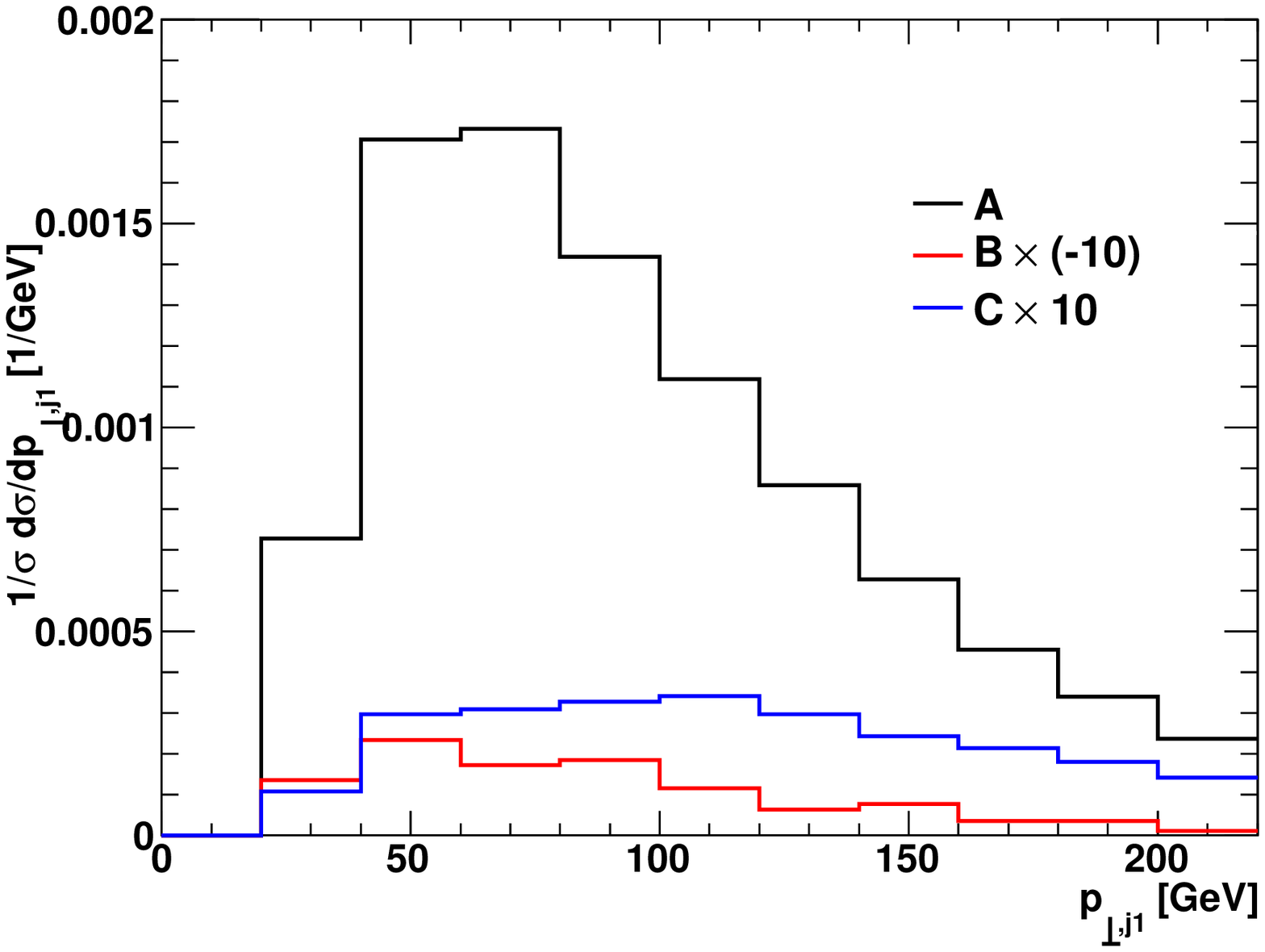} 
\caption{Contribution of the three types of virtual corrections to the
leading jet $p_{\perp}$ distribution. 
Each contribution is normalised to the next-to-leading order cross section. For a better
visibility the distributions for part B and C have been multiplied with $-10$ and $10$
respectively. For the scale we chose here $\mu=2 \, M_W$.
See text for further details.}
\label{fig:virt}
\end{center}
\end{figure}
In Fig.~\ref{fig:pt} the $p_{\perp}$ distributions of the two tagging jets are shown.
 Note that
for the leading jet, the NLO corrections lead to an increase of the cross section in the first bins
and a decrease in the high $p_{\perp}$ region. This can be explained by the possibility
of radiating a third jet at NLO, which reduces the total available energy for the two leading jets and  
hence softens the leading jet energy.

A similar behaviour can be observed for the leptons. In Fig.~\ref{fig:pte} we plot the distribution of the transverse
momentum of the electron, and one can observe that the NLO corrections slightly
shift the curve towards smaller $p_\perp$ values. 

Fig.~\ref{fig:phi} shows the distribution of the azimuthal opening angle $\phi_{e^+\mu^-}$ between the two leptons.
This distribution, relevant in the context of Higgs searches, peaks at $\phi_{e^+\mu^-} = \pi$, 
in contrast to the case of leptons coming from a
$H \to W^{+}W^{-}$ decay, where the largest contributions stem from $\phi_{e^+\mu^-} \to 0$ \cite{Rainwater:1999sd,Melia:2011dw}.

\subsection{Relevance of the different virtual contributions}
Now we briefly discuss the impact of the virtual contributions of type A, B, and C described in Section 2.
The numerical relevance of parts B and C is small: 
at the scale $\mu=2\cdot M_W$, the absolute value of part B amounts to about $~1\%$ of the size of part A 
and part C to about $~3\%$.
The contributions coming from parts B and C have been systematically included in all the results presented in this paper.

It is interesting to see whether the three types
of virtual corrections differ in their kinematics. 
In Fig.\,\ref{fig:virt} we plot the individual contribution of Part A, B, and C
to the $p_{\perp}$ distribution of the leading jet ($\mu=2 \cdot M_W$). The three parts are all normalised
to the next-to-leading order cross section at the scale $\mu=2 \cdot M_W$ in order to project out only their kinematical behaviour. 
Parts A and B show a very similar shape whereas part C peaks at higher $p_{\perp}$ values and
has a broader spectrum. As the three different parts do not necessarily have the same behaviour under a variation of the scales,  
the differences in shape will be different at a different scale. However, as the overall sizes of part B and part C are small,
we can infer that their contribution does not affect much the shape of the distributions shown in Section~\ref{sec:dist}.


\section{Conclusions}

We have calculated the NLO QCD corrections to the production of a $W^+W^-$ pair 
in association with two jets at the LHC,  
including leptonic decays of the W-bosons with full spin correlations.
The calculation was performed using the {\sc GoSam}\,\cite{Cullen:2011ac} package
for the virtual contributions, 
in combination with {\tt MadGraph/MadEvent}~\cite{Stelzer:1994ta,Maltoni:2002qb} 
and {\tt MadDipole}~\cite{Frederix:2008hu,Frederix:2010cj} for the real radiation 
and the phase space integration.
The virtual amplitude also includes contributions 
with vector bosons attached to closed fermion loops (called part B) and contributions from quarks of the 
third generation in the loops (called part C), which have not been considered in the previous literature.
{\sc GoSam} provides a setup in which these contributions can be easily filtered 
at the level of Feynman diagrams. We find that, at the scale $\mu=2\,M_W$, 
part B amounts to about $~1\%$, while part C makes up about $~3\%$ of the size of the 
virtual contributions. 
The latter part also shows a different shape in its $p_{\perp}$ distribution, 
but as it is numerically small, the impact on the full $p_{\perp}$
distribution is not significant.

Overall, we find NLO corrections of the order of 10\% for standard cuts at the LHC.
The scale dependence is considerably reduced by the NLO corrections,  
from about 30\% at LO to about 6\% at NLO. Further phenomenological studies, including also 
vector boson fusion cuts, are in progress.

Our results show that the package {\sc GoSam} for automated one-loop calculations, 
in combination with a suitable program for the real radiation part, is able to 
deal with non-trivial multi-leg calculations efficiently.


\section*{Acknowledgements}

The work of N.G. was supported in part by the U.S. Department of Energy under contract
No. DE-FG02-91ER40677.
P.M. and T.R. were supported by the Alexander von
Humboldt Foundation, in the framework of the Sofja Kovaleskaja Award Project
``Advanced Mathematical Methods for Particle Physics'', endowed by the German
Federal Ministry of Education and Research.
The work of G.O. was supported in part by the National Science Foundation
under Grant PHY-0855489 and PHY-1068550.
The research of F.T. is supported by Marie-Curie-IEF, project:
``SAMURAI-Apps''.
We also acknowledge the support of the Research Executive Agency (REA)
of the European Union under the Grant Agreement number
PITN-GA-2010-264564 (LHCPhenoNet).


\appendix

\section{Numerical results at amplitude level}

To facilitate comparisons, we use the same phase space point as in \cite{Melia:2011dw}
but with the conventions 
$$p_1+p_2\to \sum_{i=3}^8p_i\, .$$
Numerical values for the momenta $p_1 \ldots p_8$ from Ref.~\cite{Melia:2011dw} are reported in Table~\ref{tabps}.

\begin{table*}[ptb]
\begin{center}
\begin{kinematics}
$p_1$ & 500&500&0&0 \\
$p_2$ & 500&-500&0&0 \\
$p_{q/g}$ & 214.48887016141776&-27.06079802177751&-98.519808378615&188.59224795994947\\
$p_{\bar{q}/g}$ &  54.23140701179994&-31.13301620817981& -7.9279665679114&43.69128236111634 \\
$p_{\nu_e}$ & 85.5312248384887&-8.22193223977868&36.16378376820329&-77.0725048002413 \\
$p_{e^+}$ & 181.42881161004266&-57.85998294819373&-171.863734086635
&-5.611858984813\\
$p_{\mu^-}$ & 82.84930107743558&-65.90954762358915& -49.89521571962871&5.51413360058664\\
$p_{\bar{\nu}_\mu}$ & 381.47038530081545&190.18527704151887&292.042940984587&-155.113300136598
\end{kinematics}
\end{center}
\caption{Kinematic point used for all the tables in this Appendix}.
\label{tabps}
\end{table*}

The renormalization scale is $\mu_R = 150$\,GeV, and for the points 
given in this appendix we use  $M_W=80.419$ \,GeV, $\Gamma_W=2.141$ \,GeV,
$\Gamma_Z=2.49$ \,GeV, $\sin\theta_W=0.47138095$.
The numerical results for the ratios of the unrenormalised virtual amplitudes to the tree-level 
amplitudes are given in Table~\ref{subresults}. The eight partonic configurations correspond to the basic partonic processes, 
listed in Section~2.2,
from which all the others can be obtained via crossings and/or change of quark generation. 

\begin{table}[htb]  {\small
\begin{center}
\begin{tabular}{|l|r|r|r|}
\hline
 Part A & $ \eps^{-2} $ & $\eps^{-1}$ & $\eps^0$  \\
\hline
$u,\bar{u},c,\bar{c}$ &$-5.333333$& $ 7.587051$ & $5.395241$  \\
$u,\bar{u},d,\bar{d}$ &$-5.333333$  & $7.679374$  & $2.155533$ \\
$d,\bar{d},s,\bar{s}$ & $-5.333333$ &  $7.587051$ &  $7.040878$ \\
\hline 
$u,\bar{u},u,\bar{u}$ & $-5.333333$ & $6.828238$ & $-2.016698$\\
$d,\bar{d},d,\bar{d}$ & $-5.333333$ & $6.733070$ & $3.214751$ \\
\hline
$d,\bar{u},c,\bar{s}$
&$-5.333333$& $ 7.587051$ & $   -15.91575 $  \\
\hline 
$u,\bar{u},g,g$ & $-8.666666$ & $-2.786885$ & $-4.712831$ \\
$d,\bar{d},g,g$ & $-8.666666$ &  $-2.836720$ & $-0.7309734$ \\
\hline
\end{tabular} 
\end{center} }
\caption{Numerical results for the ratio of virtual over tree-level
  squared amplitudes summed over helicities and colors, $N_f=4$,
  in the same approximation as in Ref.~\cite{Melia:2011dw}, i.e. 
  {\it without } contributions from 3rd generation quarks in the loops
  and {\it without} contrubutions where vector bosons are attached to closed fermion loops. 
  Only the partons accompanying the W-boson pair are listed. }
\label{subresults}
\end{table}

For a more detailed comparison, we also provide the results for the finite parts of the unrenormalised amplitudes
in the form $a + b N_f$. The results, for all subprocesses, are contained in Table~\ref{nfresults}.

\begin{table}[htb]  {\small
\begin{center}
\begin{tabular}{|l|r|r|}
\hline
 $a + b N_f$ & $a$ & $b$   \\
\hline
$u,\bar{u},c,\bar{c}$ & $14.575491$	& $-2.295061$\\
$u,\bar{u},d,\bar{d}$ & $10.222774$ &$-2.016810$\\
$d,\bar{d},s,\bar{s}$ & $16.296114$	&$-2.313809$\\
\hline 
$u,\bar{u},u,\bar{u}$ & $5.169238$&	$-1.796484$\\
$d,\bar{d},d,\bar{d}$ & $9.441580$	&$-1.556707$\\
\hline
$d,\bar{u},c,\bar{s}$ & $-19.781800$ &	$0.966511$ \\
\hline 
$u,\bar{u},g,g$ &	$-4.660510$	&$-0.013080$\\
$d,\bar{d},g,g$ & $-0.678101$	&$-0.013218$\\
\hline
\end{tabular} 
\end{center} }
\caption{Numerical results for the ratio of virtual (finite part) over tree-level
  squared amplitudes summed over helicities and colours,
  in the same approximation as Table~\ref{subresults}. 
  To facilitate the comparison, we rewrite our results for the unrenormalised 
  amplitudes in the form $a + b N_f$.}
\label{nfresults}
\end{table}



Tables~\ref{floopsbos} and~\ref{3rdgen}  contain two classes of diagrams that have not been considered before, 
namely contributions in which vector bosons are attached to a closed fermion loops (Table~\ref{floopsbos}),
and contributions coming from the inclusion of the
third quark generation in the loops (Table~\ref{3rdgen}).
Only the partons accompanying the $W$-boson pair are listed.

\begin{table}[htb]  {\small
\begin{center}
\begin{tabular}{|l|r|r|r|}
\hline
 Part B & $ \eps^{-2} $ & $\eps^{-1}$ & $\eps^0$  \\
\hline
$u,\bar{u},c,\bar{c}$& $0$& $0$& $ 0.6750172$  \\
\hline 
$u,\bar{u},d,\bar{d}$ &$0$& $0$ & $0.4313884$\\
$d,\bar{d},s,\bar{s}$ &$0$& $0$ & $0.4577387$\\
\hline 
$u,\bar{u},u,\bar{u}$ &$0$& $0$ & $1.226365$ \\
$d,\bar{d},d,\bar{d}$ &$0$& $0$ & $-0.4237687$ \\
\hline
$d,\bar{u},c,\bar{s}$&$0$& $0$ &  $0$\\
\hline
$u,\bar{u},g,g$ &$0$& $0$ & $0.6780958$\\
$d,\bar{d},g,g$ &$0$& $0$ & $0.4948705$ \\
\hline
\end{tabular}
\end{center} }
\caption{Numerical results for the ratio of virtual over tree-level
  squared amplitudes  for diagrams with vector bosons  
  attached to fermion loops (first two generations only), summed over all helicities and colors.}
\label{floopsbos}
\end{table}

\begin{table}[htb] {\small
\begin{center}
\begin{tabular}{|l|r|r|r|}
\hline
 Part C & $ \eps^{-2} $ & $\eps^{-1}$ & $\eps^0$   \\
\hline
$u,\bar{u},c,\bar{c}$ &$0$& $-1.333333$ & $-1.734964$ \\
$u,\bar{u},d,\bar{d}$ &$0$& $-1.333333$ & $-1.641250$ \\
$d,\bar{d},s,\bar{s}$ &$0$& $-1.333333$ & $-2.197013$ \\
\hline 
$u,\bar{u},u,\bar{u}$ &$0$& $-1.333333$ & $-1.713175$ \\
$d,\bar{d},d,\bar{d}$ &$0$& $-1.333333$ & $-1.179151$ \\
\hline
$d,\bar{u},c,\bar{s}$
&$0$& $-1.333333$ & $1.654891$ \\
\hline 
$u,\bar{u},g,g$ &$0$&$0$& $0.3513208$ \\
$d,\bar{d},g,g$ &$0$&$0$& $0.2061196$ \\
\hline
\end{tabular}
\end{center} }
\caption{Numerical results for the ratio of virtual over tree-level
  squared amplitudes  for contributions
   with 3rd generation
  quarks in the loops ($m_t=171.2$ GeV, $m_b=4.7$ GeV), summed over all helicities and colours. }
\label{3rdgen}
\end{table}


\newpage



\begin{thebibliography}{10}

\bibitem{Baur:1995uv}
U.~Baur, T.~Han, and J.~Ohnemus, ``{QCD corrections and nonstandard three
  vector boson couplings in $W^{+} W^{-}$ production at hadron colliders},''
  {\em Phys.Rev.} {\bf D53} (1996) 1098--1123,
  \href{http://arXiv.org/abs/hep-ph/9507336}{{\tt hep-ph/9507336}}.

\bibitem{Campanario:2010xn}
F.~Campanario, C.~Englert, and M.~Spannowsky, ``{QCD corrections to
  non-standard WZ+jet production with leptonic decays at the LHC},'' {\em
  Phys.Rev.} {\bf D82} (2010) 054015,
  \href{http://arXiv.org/abs/1006.3090}{{\tt 1006.3090}}.

\bibitem{Ohnemus:1991kk}
J.~Ohnemus, ``{An Order $\alpha_s$ calculation of hadronic $W^{-} W^{+}$
  production},'' {\em Phys.Rev.} {\bf D44} (1991) 1403--1414.

\bibitem{Frixione:1993yp}
S.~Frixione, ``{A Next-to-leading order calculation of the cross-section for
  the production of W+ W- pairs in hadronic collisions},'' {\em Nucl.Phys.}
  {\bf B410} (1993)
280--324.

\bibitem{Accomando:2004de}
E.~Accomando, A.~Denner, and A.~Kaiser, ``{Logarithmic electroweak corrections
  to gauge-boson pair production at the LHC},'' {\em Nucl.Phys.} {\bf B706}
  (2005) 325--371,
\href{http://arXiv.org/abs/hep-ph/0409247}{{\tt hep-ph/0409247}}.

\bibitem{Campbell:1999ah}
J.~M. Campbell and R.~Ellis, ``{An Update on vector boson pair production at
  hadron colliders},'' {\em Phys.Rev.} {\bf D60} (1999) 113006,
  \href{http://arXiv.org/abs/hep-ph/9905386}{{\tt hep-ph/9905386}}.

\bibitem{Dixon:1999di}
L.~J. Dixon, Z.~Kunszt, and A.~Signer, ``{Vector boson pair production in
  hadronic collisions at order $\alpha_s$ : Lepton correlations and anomalous
  couplings},'' {\em Phys.Rev.} {\bf D60} (1999) 114037,
  \href{http://arXiv.org/abs/hep-ph/9907305}{{\tt hep-ph/9907305}}.

\bibitem{Grazzini:2005vw}
M.~Grazzini, ``{Soft-gluon effects in WW production at hadron colliders},''
  {\em JHEP} {\bf 0601} (2006) 095,
\href{http://arXiv.org/abs/hep-ph/0510337}{{\tt hep-ph/0510337}}.

\bibitem{Campbell:2009kg}
J.~Campbell, E.~Castaneda-Miranda, Y.~Fang, N.~Kauer, B.~Mellado, {\em et al.},
  ``{Normalizing Weak Boson Pair Production at the Large Hadron Collider},''
  {\em Phys.Rev.} {\bf D80} (2009) 054023,
\href{http://arXiv.org/abs/0906.2500}{{\tt 0906.2500}}.

\bibitem{Campbell:2011bn}
J.~M. Campbell, R.~Ellis, and C.~Williams, ``{Vector boson pair production at
  the LHC},'' {\em JHEP} {\bf 1107} (2011) 018,
\href{http://arXiv.org/abs/1105.0020}{{\tt 1105.0020}}.

\bibitem{Agarwal:2010sn}
N.~Agarwal, V.~Ravindran, V.~K. Tiwari, and A.~Tripathi, ``{Next-to-leading
  order QCD corrections to $W^+W^-$ production at the LHC in Randall Sundrum
  model},'' {\em Phys.Lett.} {\bf B690} (2010) 390--395,
\href{http://arXiv.org/abs/1003.5445}{{\tt 1003.5445}}.

\bibitem{Campbell:2007ev}
J.~M. Campbell, R.~Ellis, and G.~Zanderighi, ``{Next-to-leading order
  predictions for $WW+1$ jet distributions at the LHC},'' {\em JHEP} {\bf 0712}
  (2007) 056, \href{http://arXiv.org/abs/0710.1832}{{\tt 0710.1832}}.

\bibitem{Dittmaier:2009un}
S.~Dittmaier, S.~Kallweit, and P.~Uwer, ``{NLO QCD corrections to pp/ppbar
  $\to$ WW+jet+X including leptonic W-boson decays},'' {\em Nucl.Phys.} {\bf
  B826} (2010) 18--70, \href{http://arXiv.org/abs/0908.4124}{{\tt 0908.4124}}.

\bibitem{Bern:2008ef}
{\bf NLO Multileg Working Group} Collaboration, Z.~Bern {\em et al.}, ``{The
  NLO multileg working group: Summary report},''
  \href{http://arXiv.org/abs/0803.0494}{{\tt 0803.0494}}.

\bibitem{Binoth:2005ua}
T.~Binoth, M.~Ciccolini, N.~Kauer, and M.~Kramer, ``{Gluon-induced WW
  background to Higgs boson searches at the LHC},'' {\em JHEP} {\bf 0503}
  (2005) 065,
\href{http://arXiv.org/abs/hep-ph/0503094}{{\tt hep-ph/0503094}}.

\bibitem{Binoth:2006mf}
T.~Binoth, M.~Ciccolini, N.~Kauer, and M.~Kramer, ``{Gluon-induced W-boson pair
  production at the LHC},'' {\em JHEP} {\bf 0612} (2006) 046,
  \href{http://arXiv.org/abs/hep-ph/0611170}{{\tt hep-ph/0611170}}.

\bibitem{Campbell:2011cu}
J.~M. Campbell, R.~Ellis, and C.~Williams, ``{Gluon-Gluon Contributions to W+
  W- Production and Higgs Interference Effects},'' {\em JHEP} {\bf 1110} (2011)
  005,
\href{http://arXiv.org/abs/1107.5569}{{\tt 1107.5569}}.

\bibitem{Duhrssen:2005bz}
M.~Duhrssen, K.~Jakobs, J.~van~der Bij, and P.~Marquard, ``{The Process gg to
  WW as a background to the Higgs signal at the LHC},'' {\em JHEP} {\bf 0505}
  (2005) 064,
\href{http://arXiv.org/abs/hep-ph/0504006}{{\tt hep-ph/0504006}}.

\bibitem{Jager:2006zc}
B.~Jager, C.~Oleari, and D.~Zeppenfeld, ``{Next-to-leading order QCD
  corrections to $W^+ W^-$ production via vector-boson fusion},'' {\em JHEP}
  {\bf 07} (2006) 015,
\href{http://arXiv.org/abs/hep-ph/0603177}{{\tt hep-ph/0603177}}.

\bibitem{Jager:2009xx}
B.~Jager, C.~Oleari, and D.~Zeppenfeld, ``{Next-to-leading order QCD
  corrections to $W^+W^+$jj and $W^-W^-$jj production via weak-boson fusion},''
  {\em Phys. Rev.} {\bf D80} (2009) 034022,
\href{http://arXiv.org/abs/0907.0580}{{\tt 0907.0580}}.

\bibitem{Melia:2010bm}
T.~Melia, K.~Melnikov, R.~Rontsch, and G.~Zanderighi, ``{Next-to-leading order
  QCD predictions for $W^+W^+jj$ production at the LHC},'' {\em JHEP} {\bf
  1012} (2010) 053, \href{http://arXiv.org/abs/1007.5313}{{\tt 1007.5313}}.

\bibitem{Melia:2011gk}
T.~Melia, P.~Nason, R.~Rontsch, and G.~Zanderighi, ``{$W^+W^+$ plus dijet
  production in the POWHEGBOX},'' \href{http://arXiv.org/abs/1102.4846}{{\tt
  1102.4846}}.

\bibitem{Nason:2004rx}
P.~Nason, ``{A New method for combining NLO QCD with shower Monte Carlo
  algorithms},'' {\em JHEP} {\bf 0411} (2004) 040,
  \href{http://arXiv.org/abs/hep-ph/0409146}{{\tt hep-ph/0409146}}.

\bibitem{Frixione:2007vw}
S.~Frixione, P.~Nason, and C.~Oleari, ``{Matching NLO QCD computations with
  Parton Shower simulations: the POWHEG method},'' {\em JHEP} {\bf 0711} (2007)
  070, \href{http://arXiv.org/abs/0709.2092}{{\tt 0709.2092}}.

\bibitem{Jager:2011ms}
B.~Jager and G.~Zanderighi, ``{NLO corrections to electroweak and QCD
  production of $W^+W^+$ plus two jets in the POWHEGBOX},'' {\em JHEP} {\bf 1111}
  (2011) 055,
\href{http://arXiv.org/abs/1108.0864}{{\tt 1108.0864}}.

\bibitem{Melia:2011dw}
T.~Melia, K.~Melnikov, R.~Rontsch, and G.~Zanderighi, ``{NLO QCD corrections
  for $W^+W^-$ pair production in association with two jets at hadron
  colliders},'' \href{http://arXiv.org/abs/1104.2327}{{\tt 1104.2327}}.

\bibitem{Campanario:2011cs}
F.~Campanario, ``{Towards pp to VVjj at NLO QCD: Bosonic contributions to
  triple vector boson production plus jet},'' {\em JHEP} {\bf 1110} (2011) 070,
\href{http://arXiv.org/abs/1105.0920}{{\tt 1105.0920}}.

\bibitem{Berger:2010zx}
C.~Berger, Z.~Bern, L.~J. Dixon, F.~Cordero, D.~Forde, {\em et al.}, ``{Precise
  Predictions for W + 4 Jet Production at the Large Hadron Collider},'' {\em
  Phys.Rev.Lett.} {\bf 106} (2011) 092001,
  \href{http://arXiv.org/abs/1009.2338}{{\tt 1009.2338}}.

\bibitem{Ita:2011wn}
H.~Ita, Z.~Bern, L.~Dixon, F.~Febres~Cordero, D.~Kosower, {\em et al.},
  ``{Precise Predictions for Z + 4 Jets at Hadron Colliders},'' {\em Phys.Rev.}
  {\bf D85} (2012) 031501, \href{http://arXiv.org/abs/1108.2229}{{\tt
  1108.2229}}.
5 pages, 3 figures, 1 table, RevTex, corrected Z+0 jet cross section.

\bibitem{Bern:2011ep}
Z.~Bern, G.~Diana, L.~Dixon, F.~Febres~Cordero, S.~Hoeche, {\em et al.},
  ``{Four-Jet Production at the Large Hadron Collider at Next-to-Leading Order
  in QCD},''
\href{http://arXiv.org/abs/1112.3940}{{\tt 1112.3940}}.

\bibitem{Berger:2009ep}
C.~Berger, Z.~Bern, L.~J. Dixon, F.~Febres~Cordero, D.~Forde, {\em et al.},
  ``{Next-to-Leading Order QCD Predictions for W+3-Jet Distributions at Hadron
  Colliders},'' {\em Phys.Rev.} {\bf D80} (2009) 074036,
  \href{http://arXiv.org/abs/0907.1984}{{\tt 0907.1984}}.

\bibitem{Berger:2009zg}
C.~Berger, Z.~Bern, L.~J. Dixon, F.~Febres~Cordero, D.~Forde, {\em et al.},
  ``{Precise Predictions for $W$ + 3 Jet Production at Hadron Colliders},''
  {\em Phys.Rev.Lett.} {\bf 102} (2009) 222001,
  \href{http://arXiv.org/abs/0902.2760}{{\tt 0902.2760}}.

\bibitem{KeithEllis:2009bu}
R.~Ellis, K.~Melnikov, and G.~Zanderighi, ``{W+3 jet production at the
  Tevatron},'' {\em Phys.Rev.} {\bf D80} (2009) 094002,
  \href{http://arXiv.org/abs/0906.1445}{{\tt 0906.1445}}.

\bibitem{Melnikov:2009wh}
K.~Melnikov and G.~Zanderighi, ``{W+3 jet production at the LHC as a signal or
  background},'' {\em Phys.Rev.} {\bf D81} (2010) 074025,
  \href{http://arXiv.org/abs/0910.3671}{{\tt 0910.3671}}.

\bibitem{Berger:2010vm}
C.~Berger, Z.~Bern, L.~J. Dixon, F.~Cordero, D.~Forde, {\em et al.},
  ``{Next-to-Leading Order QCD Predictions for Z,$\gamma^*$+3-Jet Distributions
  at the Tevatron},'' {\em Phys.Rev.} {\bf D82} (2010) 074002,
  \href{http://arXiv.org/abs/1004.1659}{{\tt 1004.1659}}.

\bibitem{Campbell:2010cz}
J.~M. Campbell, R.~Ellis, and C.~Williams, ``{Hadronic production of a Higgs
  boson and two jets at next-to-leading order},'' {\em Phys.Rev.} {\bf D81}
  (2010) 074023, \href{http://arXiv.org/abs/1001.4495}{{\tt 1001.4495}}.

\bibitem{Bredenstein:2009aj}
A.~Bredenstein, A.~Denner, S.~Dittmaier, and S.~Pozzorini, ``{NLO QCD
  corrections to pp $\to$ t anti-t b anti-b + X at the LHC},'' {\em
  Phys.Rev.Lett.} {\bf 103} (2009) 012002,
  \href{http://arXiv.org/abs/0905.0110}{{\tt 0905.0110}}.

\bibitem{Bredenstein:2010rs}
A.~Bredenstein, A.~Denner, S.~Dittmaier, and S.~Pozzorini, ``{NLO QCD
  corrections to top anti-top bottom anti-bottom production at the LHC: 2. full
  hadronic results},'' {\em JHEP} {\bf 1003} (2010) 021,
  \href{http://arXiv.org/abs/1001.4006}{{\tt 1001.4006}}.

\bibitem{Bevilacqua:2009zn}
G.~Bevilacqua, M.~Czakon, C.~Papadopoulos, R.~Pittau, and M.~Worek, ``{Assault
  on the NLO Wishlist: pp $\to$ t anti-t b anti-b},'' {\em JHEP} {\bf 0909}
  (2009) 109, \href{http://arXiv.org/abs/0907.4723}{{\tt 0907.4723}}.

\bibitem{Bevilacqua:2010ve}
G.~Bevilacqua, M.~Czakon, C.~Papadopoulos, and M.~Worek, ``{Dominant QCD
  Backgrounds in Higgs Boson Analyses at the LHC: A Study of pp $\to$ t anti-t
  + 2 jets at Next-To-Leading Order},'' {\em Phys.Rev.Lett.} {\bf 104} (2010)
  162002, \href{http://arXiv.org/abs/1002.4009}{{\tt 1002.4009}}.

\bibitem{Binoth:2009rv}
T.~Binoth, N.~Greiner, A.~Guffanti, J.~Reuter, J.-P. Guillet, {\em et al.},
  ``{Next-to-leading order QCD corrections to pp $\to$ b anti-b b anti-b + X at
  the LHC: the quark induced case},'' {\em Phys.Lett.} {\bf B685} (2010)
  293--296, \href{http://arXiv.org/abs/0910.4379}{{\tt 0910.4379}}.

\bibitem{Greiner:2011mp}
N.~Greiner, A.~Guffanti, T.~Reiter, and J.~Reuter, ``{NLO QCD corrections to
  the production of two bottom-antibottom pairs at the LHC},''
  \href{http://arXiv.org/abs/1105.3624}{{\tt 1105.3624}}.

\bibitem{Bevilacqua:2010qb}
G.~Bevilacqua, M.~Czakon, A.~van Hameren, C.~G. Papadopoulos, and M.~Worek,
  ``{Complete off-shell effects in top quark pair hadroproduction with leptonic
  decay at next-to-leading order},'' {\em JHEP} {\bf 1102} (2011) 083,
  \href{http://arXiv.org/abs/1012.4230}{{\tt 1012.4230}}.

\bibitem{Denner:2010jp}
A.~Denner, S.~Dittmaier, S.~Kallweit, and S.~Pozzorini, ``{NLO QCD corrections
  to WWbb production at hadron colliders},'' {\em Phys.Rev.Lett.} {\bf 106}
  (2011) 052001, \href{http://arXiv.org/abs/1012.3975}{{\tt 1012.3975}}.

\bibitem{Campanario:2011ud}
F.~Campanario, C.~Englert, M.~Rauch, and D.~Zeppenfeld, ``{Precise predictions
  for $W \gamma \gamma$ +jet production at hadron colliders},'' {\em
  Phys.Lett.} {\bf B704} (2011) 515--519,
\href{http://arXiv.org/abs/1106.4009}{{\tt 1106.4009}}.

\bibitem{Frederix:2010ne}
R.~Frederix, S.~Frixione, K.~Melnikov, and G.~Zanderighi, ``{NLO QCD
  corrections to five-jet production at LEP and the extraction of
  $\alpha_s(M_Z)$},'' {\em JHEP} {\bf 1011} (2010) 050,
  \href{http://arXiv.org/abs/1008.5313}{{\tt 1008.5313}}.

\bibitem{Becker:2011vg}
S.~Becker, D.~Goetz, C.~Reuschle, C.~Schwan, and S.~Weinzierl, ``{NLO results
  for five, six and seven jets in electron-positron annihilation},'' {\em
  Phys.Rev.Lett.} {\bf 108} (2012) 032005,
  \href{http://arXiv.org/abs/1111.1733}{{\tt 1111.1733}}.
5 pages.

\bibitem{vanHameren:2009dr}
A.~van Hameren, C.~Papadopoulos, and R.~Pittau, ``{Automated one-loop
  calculations: A Proof of concept},'' {\em JHEP} {\bf 0909} (2009) 106,
  \href{http://arXiv.org/abs/0903.4665}{{\tt 0903.4665}}.

\bibitem{Hirschi:2011pa}
V.~Hirschi, R.~Frederix, S.~Frixione, M.~V. Garzelli, F.~Maltoni, {\em et al.},
  ``{Automation of one-loop QCD corrections},'' {\em JHEP} {\bf 1105} (2011)
  044, \href{http://arXiv.org/abs/1103.0621}{{\tt 1103.0621}}.

\bibitem{Bevilacqua:2011xh}
G.~Bevilacqua, M.~Czakon, M.~Garzelli, A.~van Hameren, A.~Kardos, {\em et al.},
  ``{HELAC-NLO},''
\href{http://arXiv.org/abs/1110.1499}{{\tt 1110.1499}}.

\bibitem{Mastrolia:2010nb}
P.~Mastrolia, G.~Ossola, T.~Reiter, and F.~Tramontano, ``{Scattering AMplitudes
  from Unitarity-based Reduction Algorithm at the Integrand-level},'' {\em
  JHEP} {\bf 1008} (2010) 080, \href{http://arXiv.org/abs/1006.0710}{{\tt
  1006.0710}}.

\bibitem{Cullen:2011ac}
G.~Cullen, N.~Greiner, G.~Heinrich, G.~Luisoni, P.~Mastrolia, {\em et al.},
  ``{Automated One-Loop Calculations with GoSam},''
\href{http://arXiv.org/abs/1111.2034}{{\tt 1111.2034}}.

\bibitem{Stelzer:1994ta}
T.~Stelzer and W.~Long, ``{Automatic generation of tree level helicity
  amplitudes},'' {\em Comput.Phys.Commun.} {\bf 81} (1994) 357--371,
\href{http://arXiv.org/abs/hep-ph/9401258}{{\tt hep-ph/9401258}}.

\bibitem{Frederix:2008hu}
R.~Frederix, T.~Gehrmann, and N.~Greiner, ``{Automation of the Dipole
  Subtraction Method in MadGraph/MadEvent},'' {\em JHEP} {\bf 0809} (2008) 122,
  \href{http://arXiv.org/abs/0808.2128}{{\tt 0808.2128}}.

\bibitem{Frederix:2010cj}
R.~Frederix, T.~Gehrmann, and N.~Greiner, ``{Integrated dipoles with MadDipole
  in the MadGraph framework},'' {\em JHEP} {\bf 1006} (2010) 086,
  \href{http://arXiv.org/abs/1004.2905}{{\tt 1004.2905}}.

\bibitem{Maltoni:2002qb}
F.~Maltoni and T.~Stelzer, ``{MadEvent: Automatic event generation with
  MadGraph},'' {\em JHEP} {\bf 0302} (2003) 027,
  \href{http://arXiv.org/abs/hep-ph/0208156}{{\tt hep-ph/0208156}}.

\bibitem{Alwall:2007st}
J.~Alwall, P.~Demin, S.~de~Visscher, R.~Frederix, M.~Herquet, {\em et al.},
  ``{MadGraph/MadEvent v4: The New Web Generation},'' {\em JHEP} {\bf 0709}
  (2007) 028, \href{http://arXiv.org/abs/0706.2334}{{\tt 0706.2334}}.

\bibitem{Catani:1996vz}
S.~Catani and M.~Seymour, ``{A General algorithm for calculating jet
  cross-sections in NLO QCD},'' {\em Nucl.Phys.} {\bf B485} (1997) 291--419,
\href{http://arXiv.org/abs/hep-ph/9605323}{{\tt hep-ph/9605323}}.

\bibitem{Nagy:1998bb}
Z.~Nagy and Z.~Trocsanyi, ``{Next-to-leading order calculation of four jet
  observables in electron positron annihilation},'' {\em Phys.Rev.} {\bf D59}
  (1999) 014020,
\href{http://arXiv.org/abs/hep-ph/9806317}{{\tt hep-ph/9806317}}.

\bibitem{Reiter:2009ts}
T.~Reiter, ``{Optimising Code Generation with haggies},'' {\em
  Comput.Phys.Commun.} {\bf 181} (2010) 1301--1331,
\href{http://arXiv.org/abs/0907.3714}{{\tt 0907.3714}}.

\bibitem{Cullen:2010jv}
G.~Cullen, M.~Koch-Janusz, and T.~Reiter, ``{Spinney: A Form Library for
  Helicity Spinors},'' {\em Comput.Phys.Commun.} {\bf 182} (2011) 2368--2387,
\href{http://arXiv.org/abs/1008.0803}{{\tt 1008.0803}}.

\bibitem{Ossola:2006us}
G.~Ossola, C.~G. Papadopoulos, and R.~Pittau, ``{Reducing full one-loop
  amplitudes to scalar integrals at the integrand level},'' {\em Nucl.Phys.}
  {\bf B763} (2007) 147--169, \href{http://arXiv.org/abs/hep-ph/0609007}{{\tt
  hep-ph/0609007}}.

\bibitem{Ossola:2007bb}
G.~Ossola, C.~G. Papadopoulos, and R.~Pittau, ``{Numerical evaluation of
  six-photon amplitudes},'' {\em JHEP} {\bf 0707} (2007) 085,
\href{http://arXiv.org/abs/0704.1271}{{\tt 0704.1271}}.

\bibitem{Ellis:2007br}
R.~Ellis, W.~Giele, and Z.~Kunszt, ``{A Numerical Unitarity Formalism for
  Evaluating One-Loop Amplitudes},'' {\em JHEP} {\bf 0803} (2008) 003,
\href{http://arXiv.org/abs/0708.2398}{{\tt 0708.2398}}.

\bibitem{Ossola:2008xq}
G.~Ossola, C.~G. Papadopoulos, and R.~Pittau, ``{On the Rational Terms of the
  one-loop amplitudes},'' {\em JHEP} {\bf 0805} (2008) 004,
\href{http://arXiv.org/abs/0802.1876}{{\tt 0802.1876}}.

\bibitem{Mastrolia:2008jb}
P.~Mastrolia, G.~Ossola, C.~Papadopoulos, and R.~Pittau, ``{Optimizing the
  Reduction of One-Loop Amplitudes},'' {\em JHEP} {\bf 0806} (2008) 030,
\href{http://arXiv.org/abs/0803.3964}{{\tt 0803.3964}}.

\bibitem{Heinrich:2010ax}
G.~Heinrich, G.~Ossola, T.~Reiter, and F.~Tramontano, ``{Tensorial
  Reconstruction at the Integrand Level},'' {\em JHEP} {\bf 1010} (2010) 105,
  \href{http://arXiv.org/abs/1008.2441}{{\tt 1008.2441}}.

\bibitem{Melnikov:2011ai}
K.~Melnikov and M.~Schulze, ``{Top quark spin correlations at the Tevatron and
  the LHC},'' {\em Phys.Lett.} {\bf B700} (2011) 17--20,
  \href{http://arXiv.org/abs/1103.2122}{{\tt 1103.2122}}.

\bibitem{Martin:2009iq}
A.~D. Martin, W.~J. Stirling, R.~S. Thorne, and G.~Watt, ``{Parton
  distributions for the LHC},'' {\em Eur. Phys. J.} {\bf C63} (2009) 189--285,
\href{http://arXiv.org/abs/0901.0002}{{\tt 0901.0002}}.

\bibitem{Cacciari:2005hq}
M.~Cacciari and G.~P. Salam, ``{Dispelling the $N^{3}$ myth for the $k_t$
  jet-finder},'' {\em Phys.Lett.} {\bf B641} (2006) 57--61,
  \href{http://arXiv.org/abs/hep-ph/0512210}{{\tt hep-ph/0512210}}.

\bibitem{Cacciari:2008gp}
M.~Cacciari, G.~P. Salam, and G.~Soyez, ``{The Anti-k(t) jet clustering
  algorithm},'' {\em JHEP} {\bf 0804} (2008) 063,
  \href{http://arXiv.org/abs/0802.1189}{{\tt 0802.1189}}.

\bibitem{Alekhin:2011sk}
S.~Alekhin {\em et al.}, ``{The PDF4LHC Working Group Interim Report},''
\href{http://arXiv.org/abs/1101.0536}{{\tt 1101.0536}}.

\bibitem{Rainwater:1999sd}
D.~L. Rainwater and D.~Zeppenfeld, ``{Observing $H \to W^{(*)}W^{(*)} \to e^\pm
  \mu^\mp /\!\!\!{p}_T$ in weak boson fusion with dual forward jet tagging at
  the CERN LHC},'' {\em Phys. Rev.} {\bf D60} (1999) 113004,
  \href{http://arXiv.org/abs/hep-ph/9906218}{{\tt hep-ph/9906218}}.
[Erratum-ibid.D61:099901,2000].

\end{thebibliography}

\providecommand{\href}[2]{#2}\begingroup\raggedright

\end{document}